\documentclass[%
aapm,
mph,%
amsmath,amssymb,
reprint,%
]{revtex4-2}
\usepackage{graphicx}
\usepackage{dcolumn}
\usepackage{bm}
\usepackage{soul}
\usepackage{siunitx}
\usepackage{ulem}
\usepackage{xcolor}
\usepackage{xspace}
\DeclareUnicodeCharacter{2009}{\,}

\usepackage[mathlines]{lineno}
\modulolinenumbers[5]
\begin{document}
	
	\preprint{AAPM/123-QED}
	
       
       \title[]{Plasma dechirper and lens for electron beams from  laser wakefield acceleration in a tailored density profile }

	\author{T.L. Steyn$^1$}
    \email{theunis-lodewyk.steyn@universite-paris-saclay.fr}
	\author{A. Panchal$^2$}
        \author{O. Vasilovici$^1$}
        \author{F.M. Herrmann$^3$}
        \author{S. Sch\"obel$^3$}
        \author{P. Ufer$^3$}
        \author{O. Khomyshyn$^1$}
        \author{Y.-Y. Chang$^3$}
        \author{I. Moulanier$^1$}
         \author{M. Masckala$^1$}
         \author{M. Samir$^3$}
        \author{C. Ballage$^1$}
        \author{M. LaBerge$^3$}
        \author{P. D\'esesquelles$^1$}
          \author{F. Massimo$^1$}
          \author{S. Dobosz Dufr\'enoy$^2$}
          \author{U. Schramm$^3$}
          \author{A. Irman$^3$}
          \author{B. Cros$^1$}
          \email{brigitte.cros@cnrs.fr}
	\affiliation{ $^1$ LPGP, CNRS Universit\'e Paris Saclay, 91400 Orsay , France  $^2$ Universit\'e Paris-Saclay, CEA, LIDYL, 91191 Gif sur Yvette, France }%
    	\affiliation{}
        \affiliation{$^3$ Helmholtz-Zentrum Dresden Rossendorf, Bautzner Landstra\ss e 400, 01328 Dresden, Germany }%

	\date{\today}
	
	\begin{abstract}

Achieving high-quality electron beams from laser wakefield accelerators critically relies on density tailoring to control electron dynamics during injection, acceleration, and extraction. We report on the experimental observation of electron beam acceleration and  shaping, in transverse momentum and longitudinal phase space, controlled by plasma density tailoring in a gas cell.  Electron beams with a FWHM charge of 40 pC at an energy of 190 MeV,  3.4\% energy spread and an rms divergence of 0.46 mrad, corresponding to a transverse momentum spread of 0.2 $m_e c$, have been measured. These beams have  a peak spectral brightness of up to 8 pC/MeV/mrad. Simulations using experimental parameters as input show that acceleration in the plasma plateau leads to chirped electron beams which then undergo transverse momentum spread reduction in a plasma down-ramp followed by dechirping in a 10~mm long plasma tail, leading to the measured peaked spectra. 
The comparison of experimental results with and without long plasma tail confirms this analysis.  \\

\end{abstract}
	
	\maketitle

In Laser WakeField Acceleration (LWFA) \cite{Esarey2009,Hooker2013}, a high intensity, short laser pulse (a few 10~fs level) travels through an under-dense plasma,  generating plasma waves. Under the effects of the accelerating electric field sustained by these waves,  trapped electrons can gain energies in excess of hundreds of MeV over a few millimeters\cite{Esarey2009,Hooker2013,downer_diagnostics_2018}. Due to  strong accelerating gradients, different longitudinal parts of the electron beam may experience different accelerating field amplitudes, resulting in large energy spreads.
However, small energy spread electron beams are needed for applications, and several mechanisms to decrease it have been investigated. Correlated energy spreads can be minimized during acceleration using mechanisms such as beam loading\cite{tzoufras_beam_2008,tzoufras_beam_2009}, which flattens the accelerating field~\cite{lindstrom_energy-spread_2021,Kirchen2021}; rephasing \cite{dopp_energy-chirp_2018,gustafsson_combined_2024}, which shifts the relative phase of the wakefield accelerating the electron beam to compensate for chirps; and quasi-phase-stable acceleration \cite{ke_near-gev_2021}, in which the electron beam remains in an almost stable phase during acceleration.Alternatively, post-acceleration compensation methods, such as plasma dechirpers, can be used to mitigate energy chirps. In a plasma dechirper, the wakefield driven by the front of the electron beam  causes the rear of the beam to experience a decelerating field, decreasing the overall energy spread. Plasma dechirpers for ps-time duration beams originating from RF-accelerators have been demonstrated \cite{darcy_tunable_2019,wu_phase_2019}. While schemes have been proposed to dechirp beams from an LWFA using a plasma dechirper\cite{wu_near-ideal_2019}, they have not yet been demonstrated experimentally.   

To successfully transport or utilize electron beams from LWFA, reducing their  divergence is essential~\cite{Migliorati2013,antici_laser-driven_2012,Mehrling2012,Marini2024}. This is achieved by minimizing the beam's transverse momentum spread (TMS). Passive plasma lensing is a key method to decrease this TMS, utilizing the transverse focusing forces of plasma waves to collimate the charged particle beam \cite{chen_grand_1987}. Implemented downstream of an LWFA using separate components or shaped density profiles, these lenses \cite{Kuschel2016,Thaury2015lens,gustafsson_combined_2024,Chang2023} have been observed to reduce beam divergence by a factor of 2. 
 Plasma lens
accompanied by emittance preservation is referred to as adiabatic matching~\cite{Floettmann2014,Ariniello2019,Dornmair2015,Sears2010}. In the beam driven regime, it has been theoretically shown that the electron beam driving its own wakefield can impart a focusing force on the rear part of the beam\cite{Lehe2014}.

 Here we report the first experimental demonstration of post-acceleration spectral and transverse shaping,  drastically improving the quality of a chirped beam produced by LWFA, achieving 
both transverse lensing and longitudinal dechirping within a single, tailored plasma structure. 
 A plasma density down-ramp is used to reduce adiabatically the electron beam transverse momentum spread; it is followed by a long, low-density plasma tail (LPT) where the beam drives its own wakefield, and undergoes further collimation and energy spread compression. This process ultimately yields relativistic electron beams with a low TMS 
of ${0.2 \ m_e c} $ and energy spread of $6$ MeV.

\begin{figure}[ht!]
	\centering
    \includegraphics[width=0.99\columnwidth]{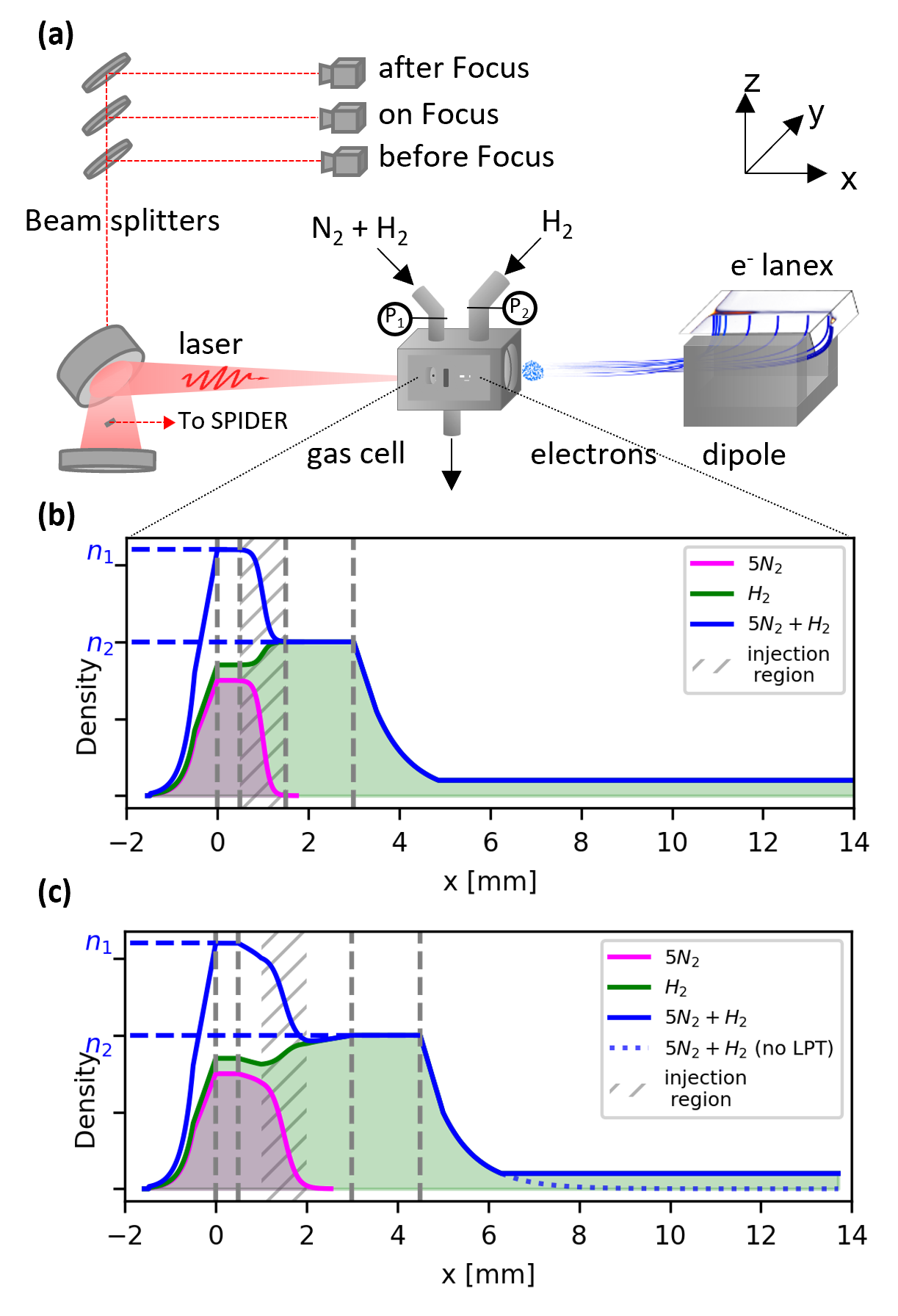} 
     \caption{Schematic of the experimental setup. The laser pulse, propagating along $x$ and linearly polarized along $y$, is focused onto a gas cell target with two compartments, which provides the desired density profile. Three cameras monitor the laser focus shot-to-shot; a SPIDER tracks the laser spectrum; and two gauges monitor shot-to-shot pressure. The focus position is defined relative to the entrance of the cell, which is at  $x=0$  mm.   In the inset, blue: electron density from hydrogen and the first five levels of nitrogen, composing the background plasma; magenta: density of the electrons from the first five levels of nitrogen. $n_1$ and $n_2$ are the plasma electron density on the density plateau of the first and second compartment respectively. Insets: density profiles along the laser propagation axis (b)  with fixed LPT for first experiment, (c)  with removable LPT and differential pumping\cite{Messner2020} between compartments for second experiment. Electron beams are characterized with a spectrometer which has a readout error of 1.4\% and 2.2 \% in divergence and energy respectively. \cite{kurz_calibration_2018,couperus_optimal_2018}}
    \label{fig:Exp_Setup}
\end{figure}

Experiments were performed using the DRACO laser at Helmholtz-Zentrum Dresden Rossendorf (HZDR), a Ti:Sa laser system with carrier wavelength $\lambda_0=0.8$ $\mu$m providing up to 2.5 J energy on target, in 30 fs FWHM pulse length with a repetition rate of up to 0.1 Hz. This laser was focused to a spot size of 24 $\mu$m FWHM into a double compartment gas cell, as illustrated in Fig.~\ref{fig:Exp_Setup}.     Shot-to-shot fluctuations from the nominal conditions were monitored using a local pressure gauge on the gas cell, spectral-phase interferometry for a direct electric-field reconstruction (SPIDER) for laser spectral-phase, and three focus-imaging cameras\cite{herrmann2026tuning} (see Supplemental Material).

We designed a gas cell with variable and replaceable 
mechanical parts to allow for shaping the density profile, as illustrated in the insets in Fig. $\ref{fig:Exp_Setup}$. The density profile used in the first experiment, with fixed LPT, is shown in Fig. $\ref{fig:Exp_Setup}$ (b) where injection and acceleration occur in the high density region ($0 <x<3.5 $ mm), which is divided into two compartments. A nitrogen-hydrogen mixture is injected into the first compartment and pure hydrogen into the second. Ionisation injection \cite{pak_injection_2010,Chen2012,Mirziae2015,Thaury2015lens,Couperus2017,Irman2018,Kurz2021,Chang2023,Picksley2024} is used to trap electrons in the wakefield, while fine control of injection location is achieved by tailoring  nitrogen density between the two compartments \cite{Pollock2011,Kirchen2021}. The two filling pressures are controlled independently, and pulsed gas injection at equal pressures ensures proper confinement~\cite{drobniak_two-chamber_2023} of the nitrogen component in the first part of the density profile. The plateau gas densities were obtained from interferometer measurements. The gas density distribution resulting in the plasma density shown in Fig. $\ref{fig:Exp_Setup}$ was derived by a parameterization of results obtained with the computational fluid dynamics code OpenFOAM \cite{weller1998,openfoam} using the multicomponentFluid package combined with the PIMPLE solver for modeling supersonic multi-component flows. 
The second density down-ramp, following the plateau of density $n_2$, is created by an $800 \text{ \textmu m}$ orifice at the outlet. This orifice leads to a cylinder of
 $4 \text{ mm}$ diameter  and $10 \text{ mm}$ length, which creates an LPT 
of almost constant low electron density ($n_{LPT} \approx 4\times 10^{16} \text{ cm}^{-3}$ for the case of Fig. $\ref{fig:Waterfalls}(a,e)$).  (See supplemental material).

For the gas cell shown in Fig. $\ref{fig:Exp_Setup}$ (b), achieving optimized electron spectra requires tuning of the laser focal position and plasma density parameters.
The optimum is found for a pressure of $20 \text{ mbar}$ in both compartments and a $15\%$ nitrogen concentration in the first, corresponding to plateau densities $n_1 = 1.6 \times 10^{18} \text{ cm}^{-3}$ and $n_2 = 1 \times 10^{18} \text{ cm}^{-3}$, with laser focus at $x = 2.4 \text{ mm}$.

The characteristics of 130 shots recorded during the first experiment for nominally constant input parameters (laser energy, focus position, gas pressure) are shown in Fig.~\ref{fig:Waterfalls} ordered by increasing mean energy: spectrum (a) and for each shot the corresponding total charge $Q_T$ (b), Median Absolute Deviation (MAD) energy spread (c), and transverse momentum spread (d).

 \begin{figure}[!htb]
 \centering
 \begin{tabular}{c}
\hspace*{-2mm}\includegraphics[width=1.03\columnwidth]{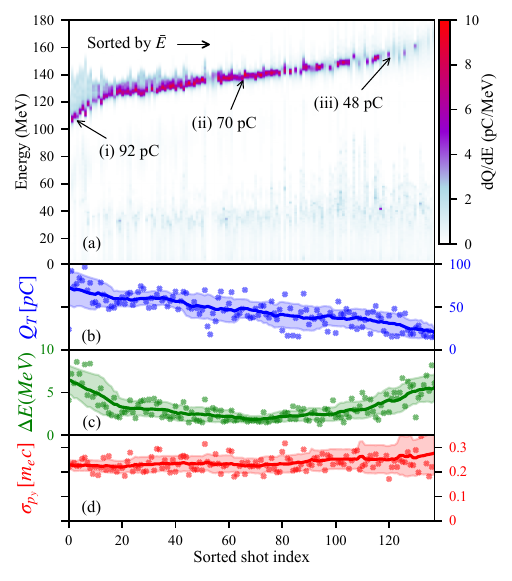}   \\[-10pt]
\hspace*{-2.4mm}\includegraphics[width=1.01\columnwidth]{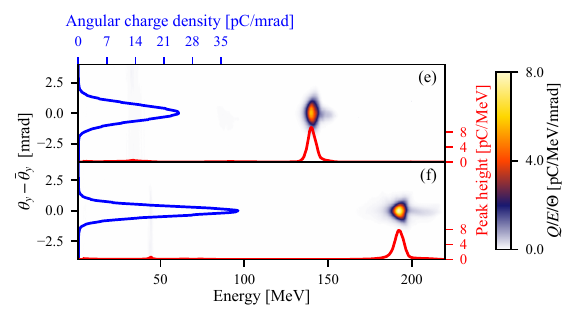}      \\
\end{tabular}
\caption{Experimental results showing the energy spectrum and energy-angle distribution of  measured electron beams. (a) Waterfall graph of energy spectra, ordered by increasing mean energy in the spectrum for a pressure of  20~mbar ($n_2 = 10^{18}$ cm$^{-3}$);  spectra labeled (i-iii)  are compared with simulations in Fig. \ref{fig:PIC_simulation_bunch_evolution2};     (b) corresponding total charge $Q_T$; blue line and shaded area are the moving average and standard deviation computed across 15 adjacent shots, respectively. (c) MAD energy spread; (d) transverse momentum spread along laser polarisation direction y; (e) and (f) are the energy-angle distribution of two electron beams obtained at respectively 20~mbar and 35 mbar ($n_2 = 1.7\times 10^{18}$ cm$^{-3}$). The pointing, $\bar{\theta}_y$, is  -0.57~mrad for (e) and -0.45 mrad for (f).
}
\label{fig:Waterfalls}
\end{figure} 


 Fig.~$\ref{fig:Waterfalls} (\text{e})$ shows the divergence-energy space of the optimal beam (ii) of Fig.~$\ref{fig:Waterfalls} (\text{a})$.
This specific shot is characterized by a peak spectral density approaching $8 \text{ pC/MeV}$ and a spectral brightness  approaching $6 \text{ pC/MeV/mrad}$.

\begin{table}[!tb]
    \caption{Summary of electron beam parameters for first experiment. [Exp] and [Sim] refer to measured and simulated parameters extracted at the end of the simulation. $E_{peak}$: energy of the spectrum peak; $\Delta E$: FWHM width of the energy spectrum peak; $Q_T$: total charge in the spectrum; $Q_{FWHM}$: charge in the FWHM peak of the energy spectrum; $\theta_{rms}$: rms divergence along the direction $y$  for the FWHM energy slice; $\sigma_{p_y}$: rms spread in the normalized transverse momentum along $y$  for the FWHM energy slice.}
    \label{table:transposed_parameters}
  \renewcommand{\arraystretch}{1.1} 
   \centering 
   \resizebox{\columnwidth}{!}{%
    \begin{tabular}{c@{\hspace{12pt}}c@{\hspace{14pt}}c@{\hspace{20pt}}c@{\hspace{20pt}}c@{\hspace{20pt}}} 
        \hline\hline 
        Parameter & Unit  & 2(e) [Exp] & 2(f) [Exp] & 3 \& 4(ii) [Sim] \\
        \hline 
        $E_{peak}$ & MeV  & $140$ & $193$ & $140$ \\
        $\Delta E$ & MeV & $5.6$ & $6.5$ & $9.0$ \\
        $Q_T$ & pC  & $70$ & $72$ & $74$ \\
        $Q_{FWHM}$ & pC & $41$ & $41$ & $41$ \\
        $\theta_{\text{rms}}$ & mrad &  $0.67$ & $0.46$ & $1.07$ \\
        $\sigma^{FWHM}_{p_y}$ & $m_ec$ &  $0.18$ & $0.17$ & $0.30$ \\
        \hline\hline 
    \end{tabular}%
    }
\end{table}




The divergence-energy space distributions for each shot are matrices $\left.dQ/d\theta_ydE\right\vert_{ij}$ in pC/MeV/mrad for each energy $E_j=\gamma_jm_ec^2$, and angle $\theta_{y,i}$, where $\gamma$ is the Lorentz factor. For ultra-relativistic beams ($\beta_x\approx1)$, $\theta_y\approx p_y/p_x$, where $p_y$ ($p_x$) is the transverse (longitudinal) momentum. At the indices $i$, $j$ the transverse momentum along $y$ is $p_{y_{i,j}}\approx\theta_{y,i}\gamma_j$. The TMS of these distributions was computed as $\sigma^2_{p_y} =\langle ( p_y-\langle p_y\rangle)^2\rangle$, where $\langle p_y\rangle$ is the average of $p_y$, calculated over the indices $ij$ using $\left.dQ/d\theta_ydE\right\vert_{ij}$ as the statistical weights. We plot the TMS determined from this discretization in Fig.~$\ref{fig:Waterfalls}(\text{d})$ for all charge above 100 MeV. The TMS is almost independent of the spectral energy. The spectra in this set have an average $\sigma_{py} = 0.23  \pm 0.03 $.

When the peak energy is increased from $140 \text{ MeV}$ (Fig. $\ref{fig:Waterfalls} (\text{e})$ ) to $190 \text{ MeV}$ (Fig. $\ref{fig:Waterfalls} (\text{f})$) the divergence is decreased from $0.67 \text{ mrad}$ to $0.46 \text{ mrad}$ while the TMS remains constant.  
This peak energy increase was achieved by increasing the density in both compartments by a factor of $1.7$ (i.e., $n_2=1.7\times 10^{18} \text{ cm}^{-3}$), which
also increased the density of the 
LPT by a factor of $1.7$. The laser focus was  adjusted to $x= 2.9 \text{ mm}$, while all other parameters remained unchanged (see supplemental material). 
The total charge and the charge in the FWHM are similar to the lower density case, as shown in Table \ref{table:transposed_parameters}. The resulting decrease of divergence leads to peaks with an increased spectral brightness of $8 \text{ pC/MeV/mrad}$ compared to the $20 \text{ mbar}$ case.


To gain insight into the physical processes leading to measured electron spectra, we performed Particle-in-Cell (PIC) simulations \cite{BirdsallLangdon2004} using the code Smilei \cite{Derouillat2018},   
with a quasi-cylindrical geometry with azimuthal Fourier decomposition \cite{Lifschitz2009}, employing two azimuthal modes.  See supplemental material for more technical details. Experimental laser and plasma parameters were used as input, together with the longitudinal hydrogen and nitrogen density profiles obtained from the parameterization of corresponding fluid simulations. The laser field distribution was modeled as a focused Flattened Gaussian Beam (FGB) \cite{Santarsiero1997} with an order $N=8$, a waist parameter of $w_0 = 14 \text{ \textmu m}$, and a Gaussian temporal profile with a pulse duration (FWHM of intensity) of $30 \text{ fs}$. This FGB fit was determined from fluence measurements taken at various positions along the laser axis around the focus \cite{Moulanier2023,Moulanier23GSAMD}. The fitting procedure provides a shot-to-shot estimate of the longitudinal focal-position jitter of between $\pm\SI{100}{\micro\meter}$ and $\pm\SI{200}{\micro\meter}$.   The normalized maximum laser electric field component along $y$, 
$a_0=\max[|E_y|]/(2\pi m_ec^2/e)\lambda_0$, was set to be $a_0=2.3$ in vacuum at the focal plane.   

In order to compare simulation and experimental results, we focus on the characteristics of the particles in the FWHM energy slice, in particular,  $\Delta E$: FWHM width of the energy spectrum peak, $Q_{FWHM}$: the amount of charge in this energy slice and  ${\sigma_{p_y}^{FWHM}}$: the transverse momentum spread of electrons within the FWHM energy slice. For the optimal shot in Fig 2 (a-ii)/(e), the comparison of the results of the simulation and experiment is given in Table \ref{table:transposed_parameters}. 

\begin{figure}[!htbp]
		\centering
			\includegraphics[width=0.99\columnwidth]{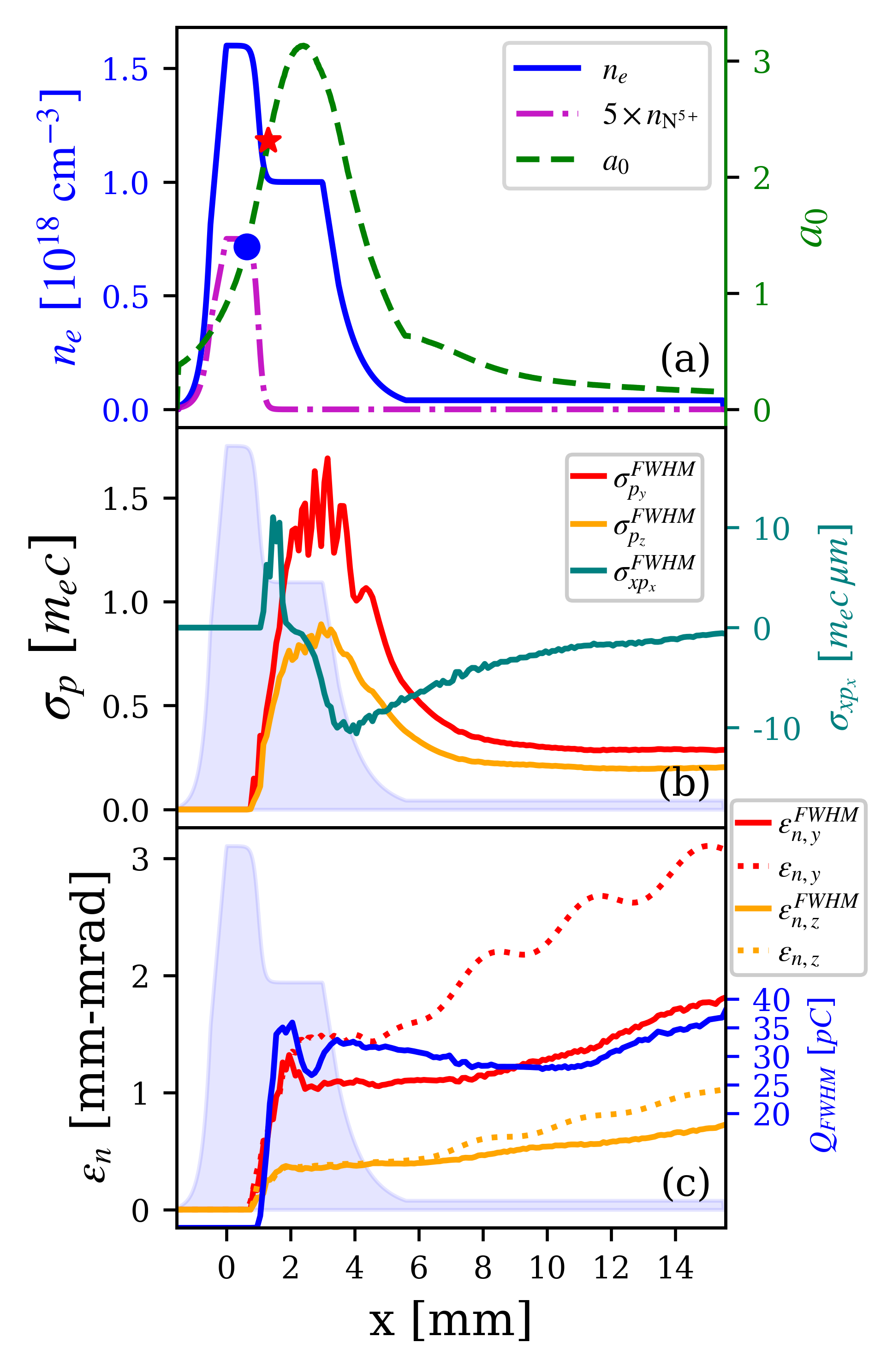}
        \caption{Simulation results showing the laser maximum electric field and electron beam evolution during the propagation in the plasma. (a) electron density profile, for all the electrons (blue) and those from the first five levels of nitrogen (magenta); the green line shows the evolution of the laser maximum normalized  transverse field $a_0$. The blue and red marker denote respectively the start and the end of the ionization of the last two levels of nitrogen. In panels (b) and (c) the total electron density profile is shown as a shaded area. (b) evolution of the TMS, perpendicular to the laser polarization direction $\sigma_{p_z}^{FWHM}$ (orange), parallel to polarization direction $\sigma_{p_y}^{FWHM}$ (red), and the longitudinal phase space correlation  $\sigma_{xp_x}^{FWHM}$ (green). (c) evolution of  normalized emittance for the $y$ (red) and $z$ (orange) axis respectively for the full beam and for the FWHM energy slice in dotted and solid lines respectively. The charge in the FWHM energy slice is plotted in blue.}
		\label{fig:PIC_simulation_bunch_evolution}
	\end{figure}

Fig. $\ref{fig:PIC_simulation_bunch_evolution}$ illustrates the laser normalized peak electric field propagation along  the plasma density distribution  (a), 
the electron beam longitudinal and transverse phase space (b), and the associated emittance (c) for the optimal shot in Fig 2 (a-ii)/(e). 


In Fig.~\ref{fig:PIC_simulation_bunch_evolution}(a), the points where the ionization of $N^{5+}$ and $N^{6+}$ starts and ends are shown by a blue disk and a red star respectively. Injection occurs within the first plasma density down-ramp of length 500 \unit{\micro\meter}. The decrease of plasma density in the first down-ramp elongates the plasma wave  and facilitates the trapping of nitrogen electrons from the 6th and 7th energy levels released by ionization \cite{Thaury2015a,Kirchen2021,Dickson2022,Marini2024}. 
The laser intensity drops rapidly in the second down-ramp ($x\simeq 4 $~mm), decreasing below the relativistic threshold ($a_0 < 1$) towards the end of the down-ramp. 
However, it remains intense enough ($a_0(x=15 \text{ mm}) = 0.2$) to ionize hydrogen gas in front of the electron beam.
The total charge $Q_T$ injected into the first plasma cavity ($74.4 \text{ pC}$) is conserved throughout acceleration in the plateau, the second down-ramp, and the LPT. 

Fig. $\ref{fig:PIC_simulation_bunch_evolution}(\text{b})$ shows the evolution of the transverse momentum spread within the FWHM energy slice in the directions parallel ($\sigma_{p_y}^{FWHM}$) and perpendicular ($\sigma_{p_z}^{FWHM}$) to the laser linear polarization direction $y$.  Betatron oscillations are clearly visible in the injection and acceleration regions. For $3 < x < 6 \text{ mm}$, the down-ramp acts as a plasma lens due to the laser-driven wakefield, which reduces $\sigma_{p_y}^{FWHM}$ to $0.5~m_e c$. For $x > 6 \text{ mm}$, electron beam dynamics are determined by the fact that the beam density exceeds the plasma density ($n_b > n_p$), and the laser intensity has significantly decreased. The simulation shows that the ratio of the longitudinal electric field driven by the beam exceeds  the laser driven one by a factor of at least 5 throughout the 
LPT. In this region ($x > 6 \text{ mm}$), $\sigma_{p_y}^{FWHM}$ is further reduced to $0.3~m_e c$.   The spread in the longitudinal phase correlation of those particles in the FWHM energy slice is defined as $ \sigma_{xp_x}^{FWHM} = \langle x p_x \rangle$. This correlation $\sigma_{xp_x}^{FWHM}$ decreases down to $-10$~$m_ec \,\mu$m  at the end of the acceleration plateau and then monotonically increasing to zero in the LPT. This is a signature of the evolution of the chirp of the electron beam, showing that dechirping in the LPT compensates the chirp acquired during acceleration. 

\begin{figure}[h]
		\centering				\includegraphics[width=0.99\columnwidth]{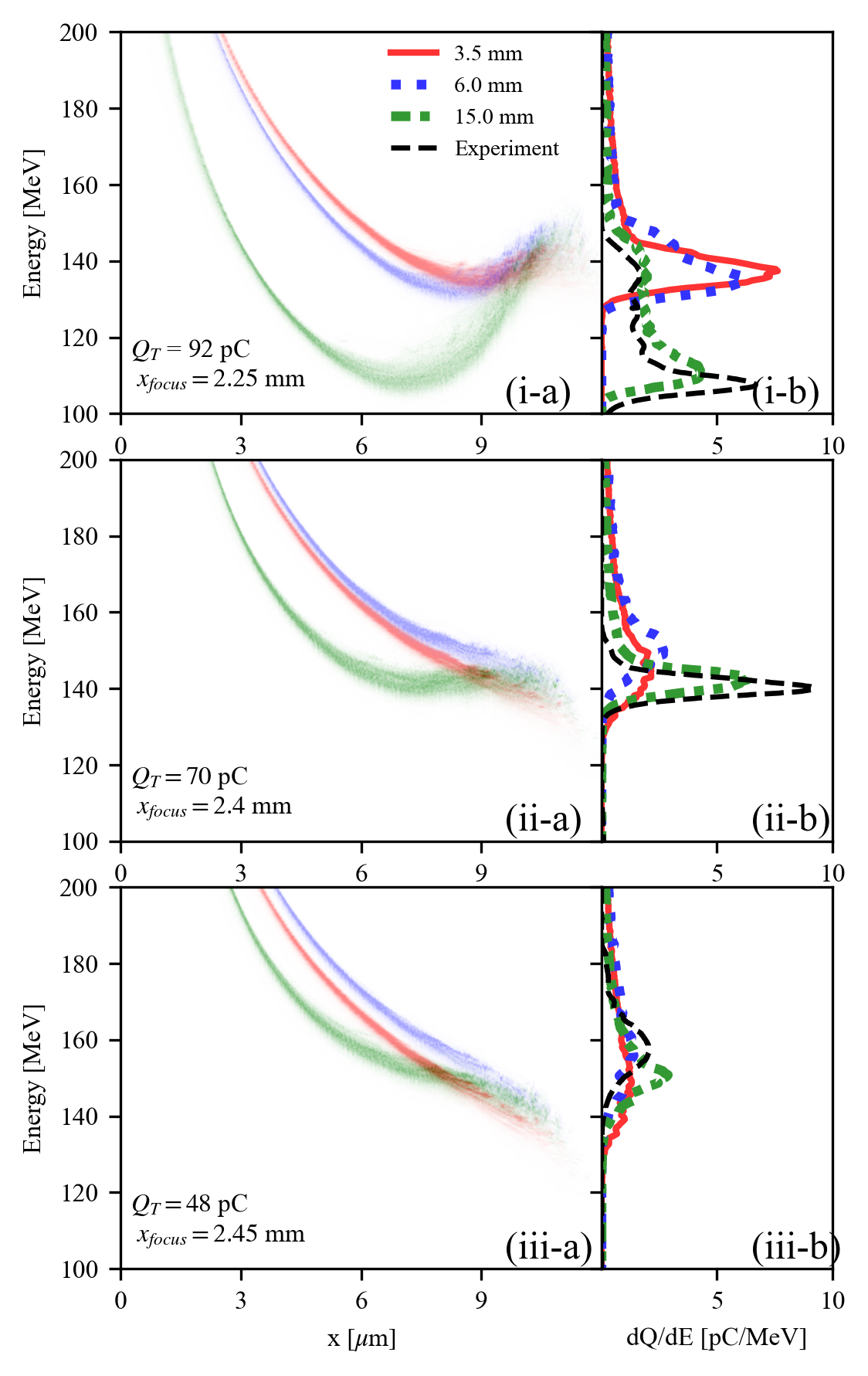} 
			\caption{Evolution of the accelerated electron beam in the plasma after the plateau. (a) Saturated longitudinal phase space distribution in the PIC simulation, at selected positions reported in the legend along density profile. 
            The longitudinal position on the horizontal axis is relative to the simulation moving window. (b) corresponding energy spectra, and experimental spectrum of the beam shown in Fig.~\ref{fig:Waterfalls}(c) (black dashed line).}
		\label{fig:PIC_simulation_bunch_evolution2}
	\end{figure}

 The full beam normalized emittance \cite{Floettmann2003} in the $k$-direction is defined as : 
\begin{equation}
    (\varepsilon_{n,k})_{rms} = \frac{1}{m_ec}\sqrt{\langle k^2\rangle \langle p_k^2 \rangle - \langle k p_k \rangle^2}
\end{equation} 
 and $\varepsilon_{n,k}^{\text{FWHM}}$ is the emittance of  the   FWHM energy slice. Here $k$ is either $y$ or $z$.
 Fig. $\ref{fig:PIC_simulation_bunch_evolution} (\text{c})$ shows the evolution of emittance  for the full beam and for those particles in the FWHM energy slice, as well as the amount of charge contained within the FWHM energy slice. The whole $74 \text{ pC}$ beam is transported through the down-ramp without emittance degradation and is thus adiabatically matched. Afterwards, the analysis of the longitudinal fields achieved in the simulations shows that beam dynamics in the LPT are governed by beam-driven wakefield, due to the drop of laser amplitude in the LPT (see Fig. $\ref{fig:PIC_simulation_bunch_evolution} (\text{a})$). As a result, electrons at the rear of the beam experience a higher focusing field than those at the front, leading to emittance growth for the full beam. In contrast, for particles in the FWHM energy slice, which are primarily located at the front of the beam (see Fig. $\ref{fig:PIC_simulation_bunch_evolution2}$),  emittance  changes in proportion to the amount of charge added to the FWHM energy slice  due to  dechirping. 

 The simulation also indicates that the transverse beam size $ \sigma_y$ ($\sigma_z$) increases from $1 \, \mu m$ (0.5) at the beginning of the down-ramp (x=3.5 mm) to $6 \, \mu m$  (4) at the end of the LPT. 

 The dechirping mechanism is illustrated in Fig. $\ref{fig:PIC_simulation_bunch_evolution2}$, showing the longitudinal phase space distribution (a) and corresponding energy spectra (b) at three longitudinal positions after the plateau from the respective simulation and the experimentally measured spectra.  The figure indices (i-iii) correspond to the the experimental spectra identified in Fig \ref{fig:Waterfalls}(a). Fig \ref{fig:PIC_simulation_bunch_evolution2}(ii) is the optimal shot corresponding to the simulation analyzed up to this point, and shown in Fig \ref{fig:Waterfalls}(c) . 

For the optimum case (ii) at $x=6 \text{ mm}$, the whole beam has a chirp of $10.9 \text{ MeV}/\text{\textmu m}$ (linear fit). Dechirping  in the LPT reduces this chirp to $7.5 \text{ MeV}/\text{\textmu m}$ over the following $9 \text{ mm}$ of propagation. This effect corresponds to a substantial dechirping strength of $380 \text{ GeV}/\text{mm}/\text{m}$ has been achieved. Considering only the particles in the final FWHM energy slice gives a similar dechirp strength, however these particles have a final residual chirp of only $0.28 \text{ MeV}/\text{\textmu m}$. This dechirping strength is two orders of magnitude larger than what has previously been reported for cases involving picosecond  beams \cite{darcy_tunable_2019}. The resulting peaked energy spectrum is in good agreement with the one experimentally measured (black dashed line). The phase space analysis confirms that the FWHM energy slice corresponds to the leading part of the bunch, which is the portion with the lowest near zero longitudinal energy chirp.

 Simulation are used to illustrate the difference between the optimum case (ii) and the suboptimal cases (i) and (iii). Although the input longitudinal laser focus position is unchanged for the 130 shots shown in Fig. 2, there is a jitter  around this nominal position is of order $\pm 100 $ \unit{\micro \meter}. Case (i) is obtained by shifting the focus in the simulation by $-150 \text{ \textmu m}$ relative to the optimum: at this focus, the laser ionizes and injects additional charge leading to a total charge of 92 pC. For case (iii) the focus is shifted by $+50 \text{ \textmu m}$, which decreases the injected charge to 48 pC. Comparing spectra for the 3 cases at x=6 mm, before any dechirping occurs, the effect of beam loading \cite{Kirchen2021} is visible. At x=6 mm,  case (i), with 90 pC, has the smallest amount of chirp, and is thus optimally beam loaded. In cases (ii) and (iii), the charge is reduced to 74~pC and 48~pC respectively, which leads to suboptimal beam loading and residual chirp at the end of the downramp. Comparing the beams at x= 15 mm shows the effect of dechirping on these beams.  The dechirping effect scales with increasing charge,
and case (i), with the largest charge, experiences the largest dechirping. However, because the beam entering the LPT had a small chirp, the resulting beam now has a negative chirp. Case (iii) has the lowest charge and as such experiences the lowest dechirp strength, and in this case it is insufficient to compensate the initial chirp. Case (ii) thus represents the optimum where simultaneously the charge and the initial chirp are  such that they can be optimally compensated in the dechirper,  leading to a narrow energy spread beam. This optimum occurs where the charge and chirp of the beam are matched to the density and length of the dechirper.  Finally, this analysis shows that the final phase space is a combination of beam loading in the accelerating section and dechirping in the LPT.

In summary, the simulations reproduce the measured beam characteristics across the observed charge range. They show that the LPT acts as both a dechirper and a lens, and that the strength of these effects depends on the total beam charge.

\begin{figure}[!ht]
 \centering
 \begin{tabular}{c}
\hspace*{-2mm}\includegraphics[width=1.03\columnwidth]{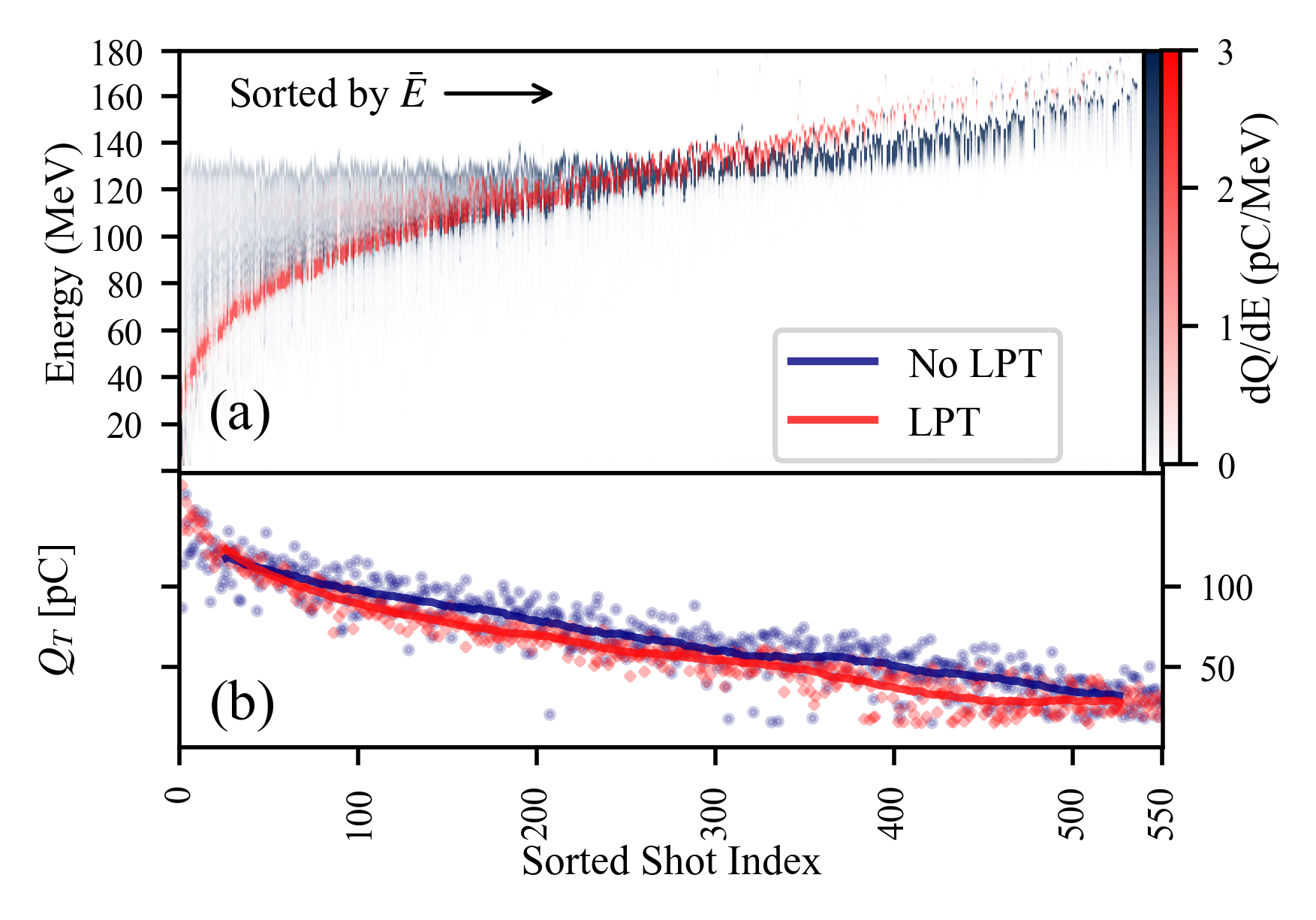}   \\  
\hspace*{-11mm}\includegraphics[width=0.97\columnwidth]{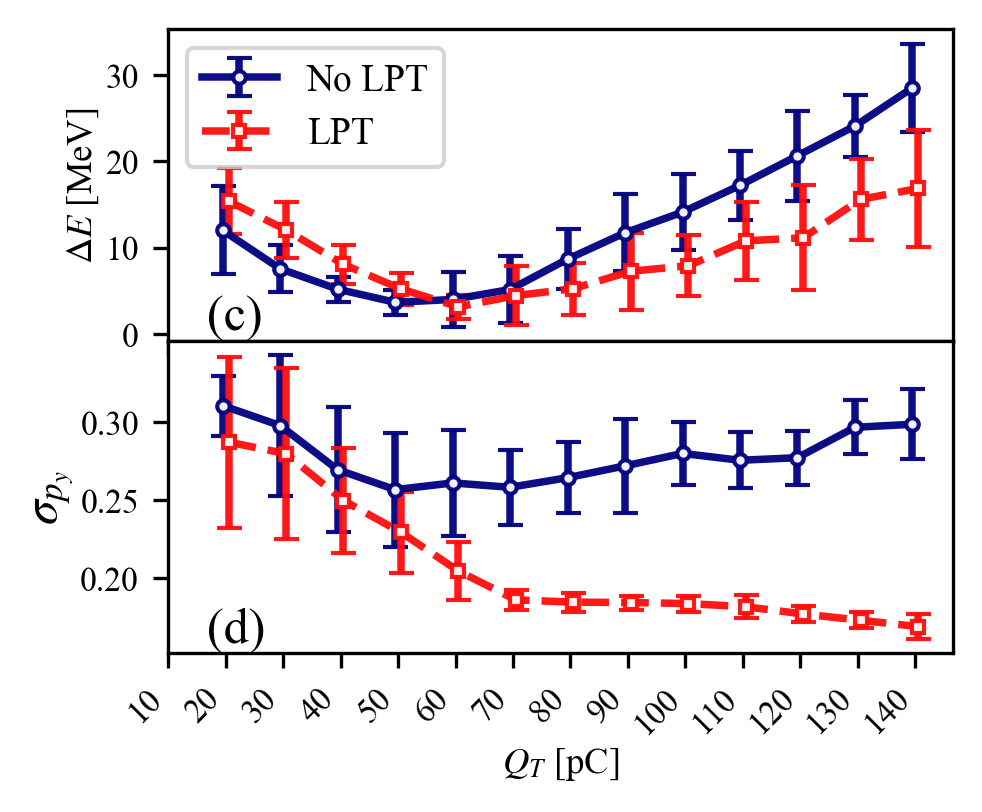}
\end{tabular}
\caption{Comparison of datasets acquired with (red) and without (blue) the long plasma tail (LPT) in the second experiment.  (a) Overlaid waterfall plot of single-shot energy spectra, shots are ordered by increasing mean energy. (b) Corresponding total charge $Q_T$ for each shot. The averages and standard deviations of (c) Median Absolute Deviation (MAD) energy spread $\Delta E$ and (d) transverse momentum spread $\sigma_{p_y}$ are computed in $Q_T$ bins of width $10\,\mathrm{pC}$ and plotted as a function of total charge; the bin labels indicate the bin centers.}
\label{fig:Waterfalls2}
\end{figure}

In order to demonstrate experimentally the  role of the LPT,  a second experiment was performed using a  gas-cell geometry enabling  the LPT section to be installed or removed, while keeping other parameters constant. The experiment was carried out with the same laser system and parameters as in the first experiment, with the same plateau densities ($n_1 = 1.6 \times 10^{18}\,\text{cm}^{-3}$ and $n_2 = 1.0 \times 10^{18}\,\text{cm}^{-3}$); when the LPT was installed, its nominal length and density were set to match the first experiment. Differential pumping \cite{Messner2020} was added between the two compartments: this improved the hydrodynamic stability of the density profile, but also introduced a change in the  density distribution at the interface between the two compartments as shown in Fig.~\ref{fig:Exp_Setup} (c).  Focus-camera measurements provided clear evidence that laser-focus fluctuations drive variations in total charge, with a Pearson correlation coefficient of up to $r=0.73$ between focus position and $Q_T$ (see Supplemental Material). The electron characteristics, with and without LPT,  are shown in Fig.~\ref{fig:Waterfalls2}.

We recorded 550 consecutive shots for each case, with and without the LPT. Figure~\ref{fig:Waterfalls2}(a) shows overlaid waterfall spectra for datasets acquired with and without the LPT, with shots ordered by increasing mean energy; Fig.~\ref{fig:Waterfalls2}(b) shows the corresponding total charge $Q_T$. To enable a like-for-like comparison, each dataset is grouped into $Q_T$ bins of width 10~pC. Figures~\ref{fig:Waterfalls2}(c) and \ref{fig:Waterfalls2}(d) show, for each bin, the mean and standard deviation of the Median Absolute Deviation (MAD) energy spread $\Delta E$ and transverse momentum spread $\sigma_{p_y}$ as a function of total charge.  $\Delta E$ is smaller in presence of  LPT at high charge. A clear reduction in $\sigma_{p_y}$ is also observed for the LPT case, particularly for $Q_T > 50$~pC. These results experimentally confirm that the long plasma tail modifies both the energy spread and the transverse momentum spread of the beam, and that these effects are the strongest at high charge, consistent with simulations.

In conclusion, reported results show that transverse lensing and longitudinal dechirping achieved within a single, tailored plasma structure, significantly improve the quality of chirped beams produced by LWFA. Experimental demonstration was achieved in a gas cell, providing a suitable plasma density distribution. The electron beam transverse momentum is reduced adiabatically in a down-ramp by the transverse laser driven wakefield.
 Lensing and energy compression by dechirping are achieved in a subsequent low density long plasma tail (LPT), where the electron beam drives its own wakefield. This process is adiabatic for the leading part of the beam and a dechirping strength of  $380 \text{ GeV}/\text{mm}/\text{m}$ has been achieved. This process yields relativistic electron beams with a low transverse momentum spread of $0.2~m_e c$, an energy spread of $6 \text{ MeV}$ and a peak brightness of 8 pC/MeV/mrad. This   demonstration of compensating the chirp of LWFA beams based on plasma tailoring  provides  an alternative path towards
the narrow energy spreads required for applications such as compact Free-Electron Lasers or multi-stage accelerator development. 

\textbf{Acknowledgements}
This project has received funding from the European Union's Horizon 2020 research and innovation programme under grant  agreement no. 871124 Laserlab-Europe. This work was granted access to the HPC
resources of TGCC and CINES under the allocations 2023-A0150510062, 2024-A0170510062 (Virtual Laplace) and 2025-A0190510062 made by GENCI. The authors acknowledge M. Bisson for the design and management of the MAITRO HPC cluster at LPGP, providing computing resources for data analysis, software development and simulations for the work presented on this article. I.~Moulanier was supported by the CNRS in the framework of the project DIANA, contract N. 1255841, and the EUPRAXIA Preparatory Phase (PP) Project, Contract No. 101079773.
The authors would like to thank T. Cloarec for his input in the early stages of this work.


\clearpage
\section*{Supplemental Material}

\subsection*{Particle in Cell Simulation}

The settings of  Particle in Cell (PIC) \cite{BirdsallLangdon2004} simulations for the results shown in Fig. 3 and 4 of the article are detailed in the following. 

Simulations were performed with the open source PIC code Smilei (\cite{Derouillat2018}), version 5.1. 

Laser and plasma parameters were varied in the range of experimental uncertainties in order to recreate the mechanisms shown in the paper. This preliminary parameter space exploration was performed using a quick laser envelope model \cite{Terzani2020,Terzani2021,Massimo2019,Massimo2019cylindrical,Massimo2020}. 

The simulations shown in the article do not use the envelope mode approximation. They were performed in quasi-cylindrical geometry with azimuthal Fourier decomposition \cite{Lifschitz2009}, using two azimuthal modes. To follow the propagation of the laser pulse, a simulation window moving along the $x$ direction at speed $c$ was used. The grid cell size is $\Delta x=$ 0.019~$\mu$m and $\Delta r=$ 0.35~$\mu$m in the longitudinal and radial directions, respectively.
The integration time-step was set to $\Delta t=$ $0.975\times dx /c = 0.063$ fs, and the simulation window size had $n_x= 3584$ and $n_r=512$ grid cells along longitudinal and radial directions, respectively.

The laser pulse, linearly polarized in the $y$ direction, was modeled with a Gaussian temporal profile with Full Width Half Maximum duration $T=30$ fs. The transverse field distribution of the simulated laser was modeled as a focused flattened Gaussian beam~\cite{Santarsiero1997} with order $N=8$, waist $w_0=$ $ 14~\mu$m and normalized peak field $a_0= \mathrm{max}[E_y]/(m_e\omega_0c/e)=2.3$, where $E_y$ is the transverse electric field along $y$ and $\omega_0=2\pi/\lambda_0$ is the carrier angular frequency, with $\lambda_0= 0.8~\mu$m. A flattened Gaussian beam is a sum of cylindrically symmetric Laguerre-Gauss modes with radial index up to $N$ and waist equal to $w_0\sqrt{N+1}$ ($N=0$ corresponds to a Gaussian beam). The mode coefficients of this sum are derived in~\cite{Santarsiero1997}. This laser model, as well as the mentioned parameters, were chosen to fit experimental measurements.

Due to the relatively low ionization potential of hydrogen and of the first levels of nitrogen, the plasma was assumed to be composed of already ionized hydrogen and nitrogen ionized up to the fifth level, as in~\cite{Kirchen2021}. Free electrons and  ions were modeled with $[1,1,8]$ macro-particles per cell distributed regularly in the $x$, $r$ directions and along the azimuthal angle interval $\theta=[0,2\pi)$.
The plasma was assumed initially uniform in the radial direction and with a density profile shown in the article Figs. 1 and 3, which was obtained from the parametrization of the density distribution calculated with OpenFOAM simulations \cite{weller1998,openfoam} of the gas cell filling. Macro-particles are advanced in  phase-space using the Boris scheme \cite{boris:proc1973}. The $N^{5+}$ ion macro-particles, assumed immobile, were subject to tunnel ionization modeled using the Ammosov, Delone, and Krainov (ADK) ionization rate \cite{ADK1986,Massimo2020}. 

The PIC simulations used the following numerical schemes in order to increase physical accuracy:
\begin{itemize}
\item Esirkepov's charge conserving current deposition scheme \cite{esirkepov:CPC2001}.
\item the Yee-like \cite{Yee1966} finite-differences solver in the time domain for Maxwell's equations with reduced numerical dispersion described in \cite{Terzani2019}, adapted to the quasi-cylindrical geometry. 
\item the B-TIS3 interpolation scheme \cite{Bourgeois2023} to reduce the effects of numerical Cherenkov radiation and the numerical artifacts due to the space-time staggering of the electromagnetic fields in Yee-like Maxwell solvers, adapted to  cylindrical geometry. 
\item Perfectly Matched Layer (PML) boundary conditions \cite{Bouchard2025} in the upper transverse boundary of the simulation window, with 30 PML cells, and Silver-M\"ulller boundary conditions \cite{He1999,baruc1993-1} for the longitudinal window borders.

\end{itemize}

\subsection*{OpenFOAM Simulation of density in the downramp and long plasma tail}

The formation of the gas density downramp and long density tail is described in the following. We simulate gas distributions, and calculate the corresponding electron densities assuming the laser intensity remains high enough along the main propagation axis to ionize the gas and form a plasma as described in the main text.

To simulate the formation of the density distribution used during the experiment, we used the Computational Fluid Dynamics (CFD) code OpenFOAM \cite{weller1998,openfoam}. To simulate  multi-component gas species, we used the module multicomponentFluid, combined with PIMPLE solver with supersonic flows allowed.  This choice has the advantage of supporting simulations with both supersonic compressible gas flows and multi-component gas mixtures.  

The density downramp and long plasma tail are important geometric characteristics of the density profile where electron beam lensing and dechirping occur. By shaping the geometry of the wall of the gas cell outlet instead of immediately allowing the gas to escape to vacuum,  density structures leading to the formation of the long plasma tail can be created.

\begin{figure*}[!htb]
	\centering
	\begin{tabular}{c}
         \includegraphics[width=0.8\textwidth]{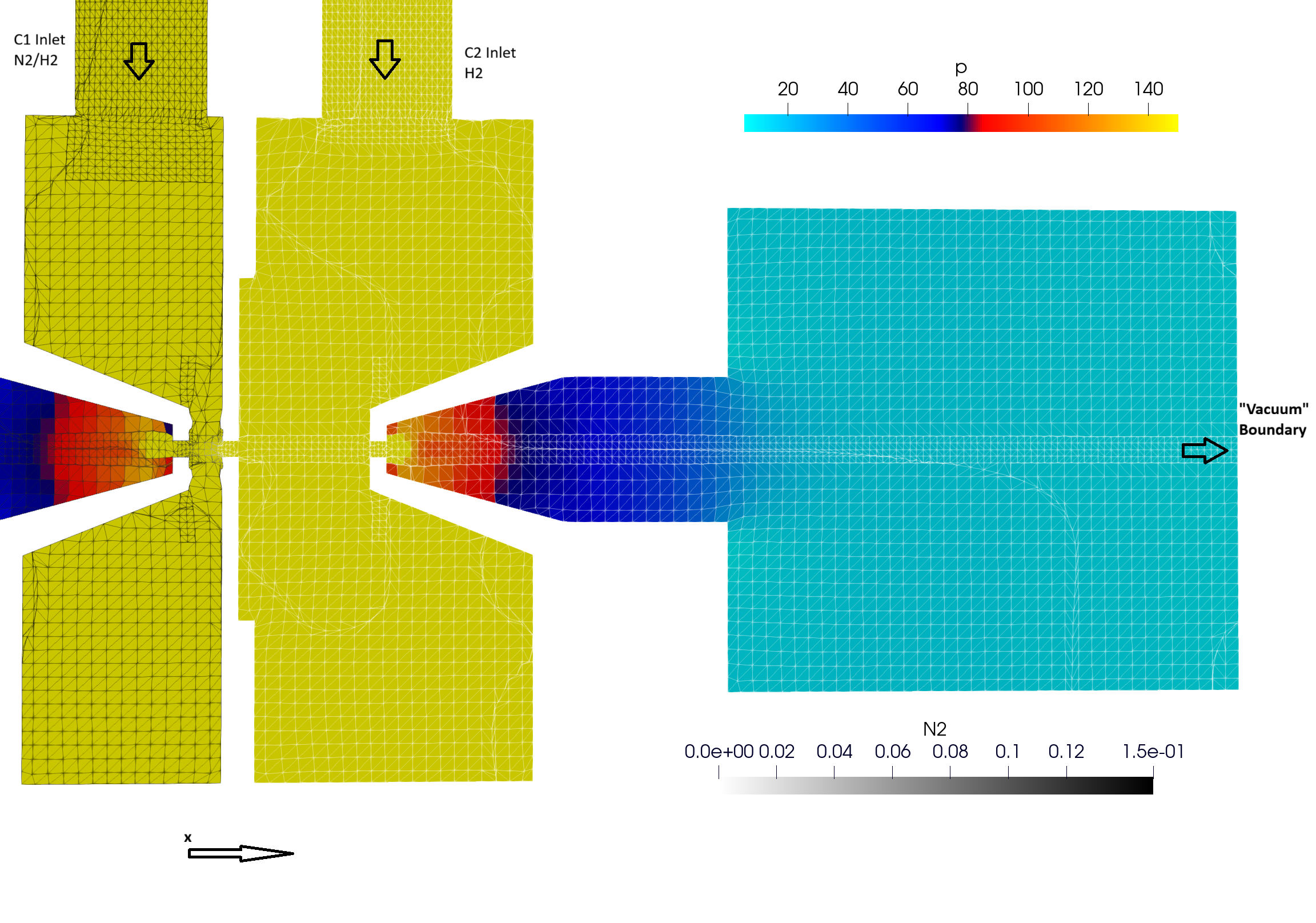} \\
	\end{tabular}
	\caption{Snapshot of OpenFOAM simulation at steady state showing the nitrogen fraction as the color of the mesh lines (black and white), and the pressure (Pa) of the gas in the compartments as the color scale. The maximum value of the colorbar is set such that the low pressure gradients become visible. Here the presence of the exponential decay after the aperture and the almost constant pressure, leading to low density plasma tail, is visible.}
	\label{fig:Openfoam}
\end{figure*}

A snapshot of the OpenFOAM simulation at steady state is shown in figure~\ref{fig:Openfoam}. The grid lines form the mesh of the simulation, and the color of the mesh (White to black) gives the concentration of nitrogen. The colorbar scale indicates the pressure in Pascal, and this colorbar is intentionally saturated at a maximum of 150 Pa (1.5 mbar) for this simulation case to highlight the areas of low density gas, in order to illustrate the location where the low density tail is formed.\\

 The right boundary condition is indicated using the  ``Vacuum boundary", where the gas reaches the ``vacuum" level. This boundary condition cannot be set to the vacuum pressure value of the experimental chamber of $10^{-5} $~mbar, as this would violate the fluid description assumptions of a continuous medium. It is critical that this boundary condition is set at a level below the pressure expected in the long plasma tail. If this is not done, the pressure will back-propagate from the boundary and the long plasma tail will not be resolved. For the simulation pictured, we set this boundary value at the level  P=10 Pa (0.1 mbar). The results of this simulation were confirmed by rerunning the simulation with this boundary condition at a lower value, P= 1 Pa (0.01 mbar). 

\begin{figure*}[!htb]
	\centering
	\begin{tabular}{c}
		 \includegraphics[width=0.55\textwidth]{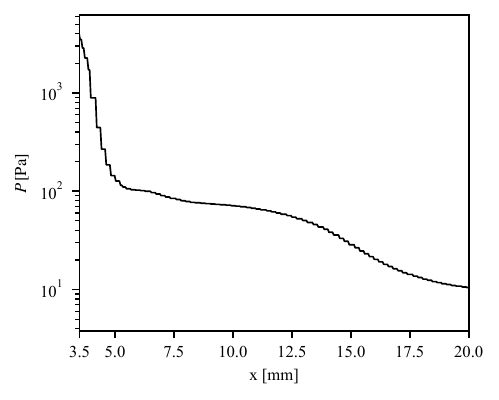} \\
	\end{tabular}
	\caption{Pressure (Pa) in the downramp and long plasma tail along the central laser axis (x) with the same (x=0 mm) reference as used in the main text. $1$ mbar = 100 Pa. }
	\label{fig:Openfoam2}
\end{figure*}

 The pressure on the laser axis in the downramp and the long plasma tail are plotted in Fig.~\ref{fig:Openfoam2}. The length along the laser axis is set with the same convention as in the main text.

From figure~\ref{fig:Openfoam2}, the exponential decay after $x = 3.5$~mm is visible as in the main text in figure 1 and figure 3. After the downramp, which in the geometry of this simulation ends at 5mm, the LPT forms. The pressure in the LPT stays almost constant between 5mm and 12.5 mm in the geometry of this simulation, changing from 100 Pa to 80 Pa between 5mm and 12.5 mm.\\
The formation of the LPT, its geometric length and the density of the gas can thus be directly obtained from the CFD model. A simplified, parameterized version of this profile is used in the Smilei simulations presented in the main text: the LPT is modeled as a constant pressure value of 0.8 mbar ($n_e = 0.04 \times 10^{18} \, cm^{-3}$ ) over 10 mm. \\

\subsection*{Measurement of the shot-to-shot laser focus fluctuations and correlation with the total charge}
\begin{figure*}[!htb]
	\centering
	\begin{tabular}{c}
		 \includegraphics[width=0.9\textwidth]{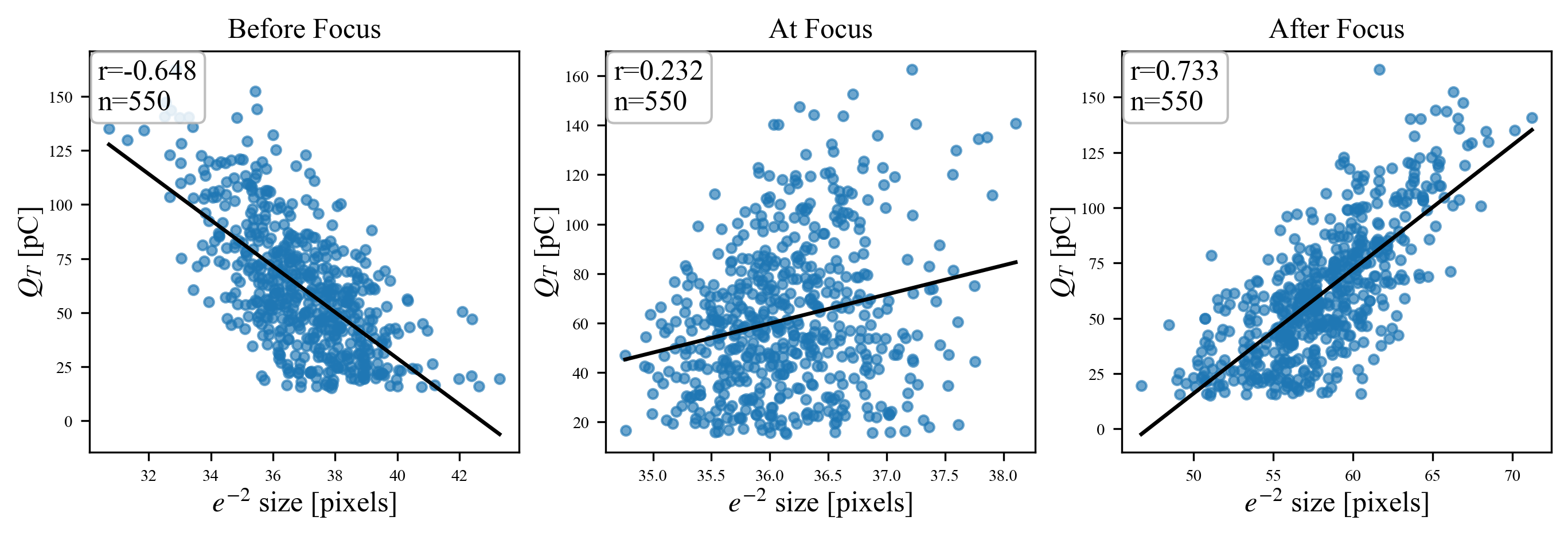} \\
	\end{tabular}
	\caption{Correlation between shot-to-shot laser focus fluctuations and total charge of the LPT case shown in Figure 5 of the main text. The $1/e^2$ size of the laser focus and the total charge for each of the three focus cameras is shown. The first plot shows the before focus camera, the second plot shows the at focus camera and the third plot shows the after focus camera. The Pearson correlation coefficient $r$ and the number of points $n$ are given for each plot.}
	\label{fig:FocusChargeCorrelation}
\end{figure*}

The shot-to-shot fluctuations of the laser focus were measured  using the three focus imaging cameras shown in the main text  figure 1. These cameras image a small fraction of the laser energy. They are positioned before the focus, at the focus and after the focus. The greatest correlation is found between the total charge and the average radius at $1/e^2$ of the recorded laser distribution on these cameras. The correlation of the 1/$e^2$ radius with the total charge is shown in figure \ref{fig:FocusChargeCorrelation}. Specifically, this dataset corresponds to the LPT case shown in figure 5 of the main text.

The figure also shows  linear fits and Pearson correlation coefficients. For the before-focus camera, a negative correlation between total charge  and radius is measured, while a positive correlation with total charge is measured after focus. There is further almost no fluctuation in the $1/e^2$ size ($\pm 3$ pixel range) at focus camera and very small correlation with the total charge. The simultaneous measurement of negative correlations before focus and positive correlations after focus is consistent with  focus position fluctuations. The measurement of the before focus $1/e^2$ size is inversely correlated with the after focus  $1/e^2$ size, with a correlation coefficient of $r= 0.70$, again consistent with  focus position fluctuations.\\

\subsection*{Waterfall plot of 35 mbar case.}

The full dataset for the 35 mbar case for which Fig 2 (f) in the main text is the optimum spectrum (Minimal energy spread) is given below in figure \ref{fig:Waterfalls_supplement}. It shows the same general trend as Fig 2 (a-d) in the 20 mbar case.

 \begin{figure}[!htb]
 \centering
 \begin{tabular}{c}
\hspace*{-2mm}\includegraphics[width=1\columnwidth]{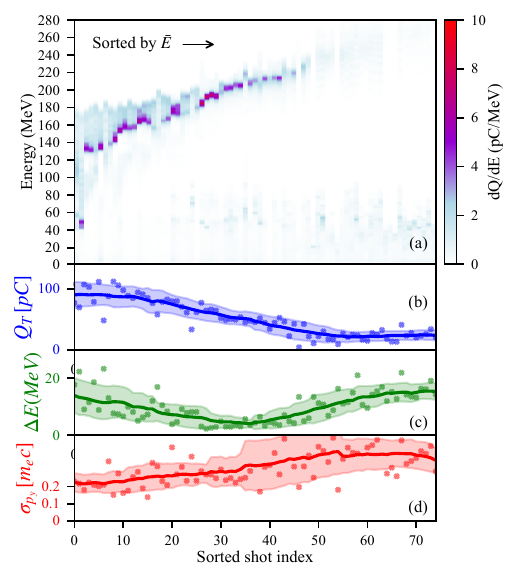}   \\[-10pt]
\end{tabular}
\caption{ Waterfall plot of electron spectra, Total charge $Q_T$ and energy spread $\Delta E$ Median Absolute Deviation (MAD) for the 35mbar case. The dataset is ordered by increasing mean energy. Fig 2 (f) is the optimum (Minimum energy spread) shot for this case.
}
\label{fig:Waterfalls_supplement}
\end{figure}

\newpage
\pagebreak[4]

\bibliographystyle{unsrt}
\bibliography{Bibliography_arxiv_merged}

\end{document}